\documentclass[oldversion]{aa}
\usepackage{graphicx}
\usepackage{deluxetable}
\usepackage[T1]{fontenc}
\usepackage{txfonts}
\usepackage{natbib}
\bibpunct{(}{)}{;}{a}{}{,}

\def\msun{{\,M_{\odot}}}

\newcommand{\der}[2]{\ensuremath{\frac{{\rm d} #1}{{\rm d} #2}}}
\newcommand{\derln}[2]{\ensuremath{\frac{{\rm dln} #1}{{\rm dln} #2}}}

\newcommand{\pderln}[2]{\ensuremath{\frac{\partial\,\rm ln\,#1}{\partial\,\rm ln\,#2}}}

\newcommand{\be}{\begin{equation}}
\newcommand{\ee}{\end{equation}}
\newcommand{\bea}{\begin{eqnarray}}
\newcommand{\eea}{\end{eqnarray}}

\makeatother
\begin{document}
    \title{Vertical dissipation profiles and the photosphere location in thin
and slim accretion disks}
   \author{
              Aleksander S\k{a}dowski\inst{1}
          \and
              Marek A. Abramowicz \inst{2,1}          
          \and \\
              Michal Bursa \inst{3}
          \and
              Wlodek Klu{\'z}niak \inst{4,1}
          \and
              Agata R{\'o}{\.z}a{\'n}ska \inst{1}
          }
   \institute{
             N. Copernicus Astronomical Center, Polish Academy
             of Sciences,
             Bartycka 18, 00-716 Warszawa,
             Poland \\
             \email{as@camk.edu.pl}
             \email{wlodek@camk.edu.pl}
             \email{agata@camk.edu.pl}
         \and
             Department of Physics, G\"oteborg University,
             SE-412-96 G\"oteborg, Sweden    \\
             \email{Marek.Abramowicz@physics.gu.se}         
         \and
            Astronomical Institute, Academy of Sciences of the
            Czech Republic,
            Bo{\v c}ni II/1401a, 141-31  Prague,
            Czech Republic
            \\
            \email{bursa@astro.cas.cz}
         \and
            Johannes Kepler Institute of Astronomy,
            Zielona G\'ora University,
            Lubuska 2, 65-265 Zielona G\'ora, Poland
             }
\date{Received ????; accepted ???? }
  \abstract{
As several authors in the past, we calculate optically thick but
geometrically thin (and slim) accretion disk models and perform a
ray-tracing of photons (in the Kerr geometry) to calculate the
observed disk spectra. Previously, it was a common practice to
ray-trace photons assuming that they are emitted from the Kerr
geometry equatorial plane, $z = 0$. We show that the spectra
calculated with this assumption differ from these calculated under
the assumption that photons are emitted from the actual surface of
the disc, $z = H(r)$. This implies that a knowledge of the
location of the thin disks effective photosphere is relevant for
calculating the spectra. In this paper we investigate, in terms of
a simple toy model, a possible influence of the (unknown, and
therefore ad hoc assumed) vertical dissipation profiles on the
vertical structure of the disk and thus on the location of the
effective photosphere, and on the observed spectra. For disks with
moderate and high mass accretion rates ($\dot m>0.01\dot m_C$) we
find that the photosphere location in the inner, radiation
pressure dominated, disk region (where most of the radiation comes
from) does not depend on the dissipation profile and therefore
emerging disk spectra are insensitive to the choice of the
dissipation function. For lower accretion rates the photosphere
location depends on the assumed vertical dissipation profile down
to the disk inner edge, but the dependence is very weak and thus
of minor importance. We conclude that the spectra of optically
thick accretion disks around black holes should be calculated with
the ray-tracing from the effective photosphere and that,
fortunately, the choice of a particular vertical dissipation
profile does not substantially influence the calculated spectrum.
}
\authorrunning{A. S\k{a}dowski et al.}
\titlerunning{Accretion disk photosphere}
  \maketitle


\section{Introduction}
\label{introduction}

There is a consensus that the observed spectra of black hole
binaries and active galactic nuclei should be explained by
accretion of rotating matter onto black holes. However, the
available theoretical models of accretion disks do not provide a
sufficiently detailed and accurate description of all physical
processes that are relevant. There is neither a quantitative
description of the turbulent dissipation that generates entropy
and transports angular momentum, nor a fully self-consistent
method to deal with the radiative transfer in the accreted matter.

Only partial solutions exist. Magnetohydrodynamic simulations
provide useful information on the ``radial''
\citep{hawleykrolik2002} and ``vertical'' \citep{turner2004}
turbulent energy dissipation, but cannot yet simultaneously deal
with the radiative transfer. On the other hand, all existing
methods of solving the radiative transfer adopt some simplifying
assumptions. Usually, they divide the flow into separate rings of
gas and assume that each ring is in a hydrostatic equilibrium.
They also assume ad hoc energy dissipation law
\citep[e.g.][]{davisblaes04,agatamadej08,idanlasota08}. In most cases,
radiative transfer with all important absorption and scattering
processes treatment is implemented on the disc surface, while the
deepest parts of the disk are treated in the diffusion
approximation. This simplification leads to a problem: the
position of the disk photosphere is calculated only very roughly
or is not calculated at all. In calculating the spectra, one
usually assumes that all the emission takes place at the
equatorial plane. However, in principle ray-tracing routines
\citep[e.g.][]{bursa.raytracing} should account for the precise
location of the emission, especially for moderate and high
accretion rates.

The spectrum computed ``vertically'' for each ring (i.e. at each
radius separately), has to be integrated over the disk surface.
This calls for a need of knowing the global ``radial'' structure
of the accretion disk, consistent with the rings ``vertical''
structures adopted in the radiative transfer calculations.
Expanding along these lines, we are working on constructing a new
\textit{vertical-plus-radial} code that inherits the sophisticated
treatment of the radiative transfer from the works mentioned
above, and at the same time fully incorporates our new radial,
fully relativistic (in the Kerr geometry), transonic slim disk\footnote{\textit{Slim disk} is
a transsonic solution of an optically thick accretion disk
described by vertically integrated equations with simplified
(limited to vertical force balance at the surface) treatment of
its vertical structure \citep{slim}.}
code \citep[described in][]{sadowski.slim}.

This paper addresses two questions concerning the effective
photosphere of optically thick and geometrically thin or slim
black hole (BH) accretion disk:

\par \textit{ Is a knowledge of the exact photosphere location
relevant for calculating the disk spectra?}

\par \textit{ Is the photosphere
location sensitive to details of dissipative processes?}


\section{The calculated spectra depend on the location of the
effective photosphere}
\label{s.spectraslim}

We start with a simple demonstration that the calculated thin disk
spectra depend on the location of the effective photosphere. For
this purpose, we use the slim disk solutions that have been
recently recalculated in the Kerr geometry by
\cite{sadowski.slim}. We choose three particular models with the
accretion rates,
\be
0.001\dot m_C, ~~0.3\dot m_C, ~~0.9\dot m_C. \label{3.rates}
\ee
Here the critical accretion rate $\dot m_C$, defined as
\be
\dot m_C=\frac{64\pi GM}{c\kappa_{es}}=2.23\times
10^{18}\frac{M}{\msun}\rm g\cdot s^{-1},
\ee
corresponds to the Eddington luminosity of an accretion disk for a
non-rotating BH.

\begin{figure}
  \resizebox{\hsize}{!}{\includegraphics{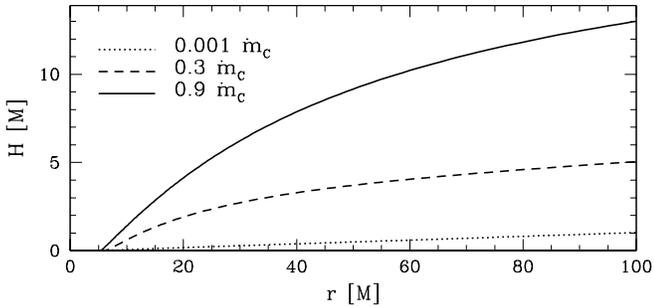}}
\caption{ Disk thickness obtained using slim disk model for
non-rotating BH. Solutions for three mass accretion rates are
presented: $\dot m = 0.001$ (dotted line), $0.3$ (dashed line) and
$0.9 \dot m_C$ (solid line). The critical mass accretion rate
$\dot m_C$ corresponds to the Eddington luminosity of an accretion
disk. } \label{f.Hslim}
\end{figure}

Disk shapes, i.e. the functions $z = H(r)$ describing the vertical
half-thickness, are presented in Fig. \ref{f.Hslim}. For the
lowest mass accretion rate ($0.001\dot m_C$) the $H/r$ ratio for
large radii is about $0.01$ while for $\dot m=0.9\dot m_C$ it
reaches $0.13$. Thus, for the accretion rates considered here,
disks are always geometrically very thin,
\begin{equation}
 H(r)\ll r.
\label{vs.hr}
\end{equation}
(Note that for larger accretion rates, the slim disk thickness
could be considerably higher.)
The models calculated by \cite{sadowski.slim} provide the local
flux of radiation $F(r, \theta_{\rm e})$ in the frame of an
observer comoving with the disk. In calculating the flux, the
standard assumption have been adopted: ({\it i}) radiation from
the effective photosphere is locally described by the black-body,
({\it ii}) the flux is limb-darkened by $F_{\rm
out}\!\propto\!2+3\cos{\theta_{\rm e}}$, where $\theta_{\rm e}$ is
the emission angle relative to the effective photosphere normal vector.
Using these initial fluxes, we ray-trace photons from the
effective photosphere to an observer located $10\rm kpc$ away from
the central black hole with mass $M_{BH}=9.4\msun$ 
(the mass corresponds to the microquasar 4U 1543-17 while the order of magnitude of the distance can be considered typical for all BH binaries) 
and compute the 
observed spectra. We take into account all
relativistic effects in the Kerr geometry
\citep{bursa.raytracing}.

For the three accretion rates (\ref{3.rates}), we calculate the
disk spectrum twice: assuming that the effective photosphere
coincides with the actual disk surface $z = H(r)$, or that it
coincides with the equatorial plane $z=0$, as if the disk would be
infinitesimally thin. This means that in the first case we start
ray-tracing from $z = H(r)$, and in the second case from $z = 0$.
\begin{figure}
  \resizebox{\hsize}{!}{\includegraphics{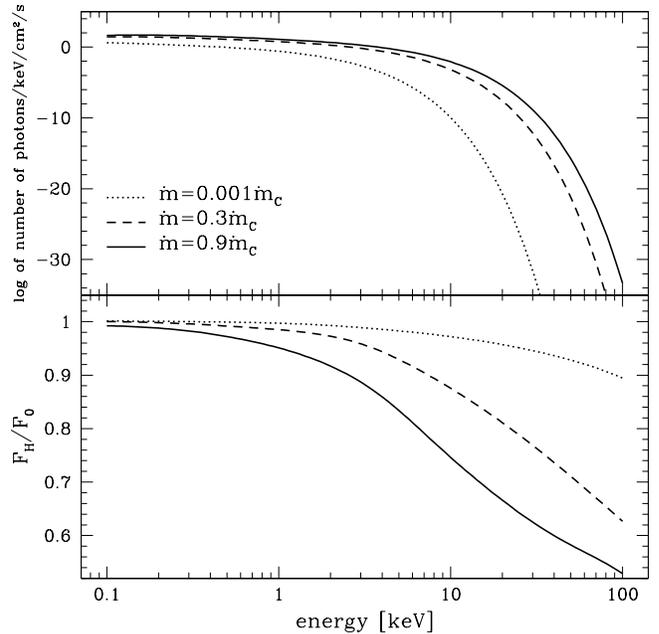}}
\caption{ Emerging spectra of accretion disks calculated using the
slim disk model for three mass accretion rates and different
assumptions about location of the emission for $M_{BH}=9.4\msun$.
The upper panel presents spectral fluxes calculated from the
photosphere assuming distance to the observer $d=10\rm kpc$ and
the inclination angle $i=60^\circ$. The ratio of the photon flux
obtained assuming photosphere emission to the photon flux obtained
from the equatorial plane is plotted in the lower panel. For the
spectral fitting purposes energy range $2\div 20 \rm keV$ is of
the major importance. } \label{f.spectraslim}
\end{figure}

The resulting spectra are plotted in Fig. \ref{f.spectraslim}.
There are obvious differences between the spectra calculated with
the two different assumptions about the location of the effective
photosphere, i.e. the photosphere either at $z = H(r)$, or at $z =
0$. Even for the moderate mass accretion rate $0.3\dot m_C$, the
``$z = 0$'' spectral flux differs from the ``$z = H(r)$'' flux by
about $20\%$ at $20 \rm keV$, and for higher accretion rates the
difference is much higher. Such differences should certainly be
considered as highly significant.

{\it Thus, our result demonstrates that knowing the precise
location of the effective photosphere is necessary in calculating
the BH slim disk spectra.} The effective photosphere lies
certainly somewhere between the $z = H(r)$ and $z = 0$ surfaces.
In principle, its location could strongly depend on details of the
vertical dissipation, and this would be a rather bad news ---
these details are still not sufficiently well known and therefore
ad hoc assumed in models.

In the rest of this paper we argue that such a situation is
unlikely. Using a simple toy model for the vertical structure of
geometrically thin, optically thick accretion disk, we show that
the photosphere location is not highly sensitive to dissipation.
Note, that for very thin (\ref{vs.hr}) stationary disks the same
is true for the total flux $F(r)$ emitted from a particular radial
location --- it does not depend on dissipation processes
\citep{shakura-73}.

These results strengthen one's confidence in estimating the black
hole spin by fitting the observed spectra of black hole sources to
these calculated theoretically \citep[see
e.g.][]{shafee-06,donegrs1915,lmcx1spin}. However, they also
indicate that the fitting procedures should be improved to include
ray-tracing from the actual location of the photosphere.

\section{A simple model of the vertical structure}
\label{sect.model}

Our simple model assumes that the accretion disk is geometrically
thin (\ref{vs.hr}) and that one may consider radial and vertical
disk structure separately.
\subsection{Radial equations}
\label{sect.radialequations}
We do not solve radial equations, assuming instead that the
rotation is strictly Keplerian (in the Kerr geometry),
\be \Omega = \Omega_K = \pm \frac{\sqrt{GM}}{r^{3/2} \pm a
\sqrt{GM/c^2}} = \sqrt{\frac{GM}{r^3}}\frac{1}{\cal B}
\label{kepler-kerr}\ee
and that the flux $F(r)$ follows from the mass, energy and angular
momentum conservation (does not depend on radial dissipation) and
is given by the standard formula \citep{nt, pagethorne},
\be F(r) = \frac{3}{8}\frac{GM{\dot M}}{\pi r^3}\frac{\cal
Q}{{\cal B}{\cal C}^{1/2}}. \label{flux-kerr}\ee
Herein $M$ is the black hole mass, $a$ is its specific angular
momentum, and
${\cal A}(r, a)$, ${\cal B}(r, a)$, ${\cal C}(r, a)$,
${\cal D}(r, a)$, ${\cal E}(r, a)$, ${\cal Q}(r, a)$
are relativistic correction
factors, defined as explicit functions of their arguments in
\citet{pagethorne}.
\subsection{Vertical equations}
\label{sect.equations}
We describe the vertical structure of the BH accretion disk in
optically thick regime by the following equations:

(i) The hydrostatic equilibrium:
\begin{equation}
 \der Pz=-\rho \Omega_K^2{\cal G}z
\end{equation}
where $P$ is the sum of the gas and radiation pressures:
\begin{equation}
 P=P_{gas}+P_{rad}=k\rho T+\frac13aT^4
\label{vs.Pressure}
\end{equation}
and $\Omega_K^2{\cal G}z$ being the vertical component of the
gravitational force of the central object calculated using the
relativistic correction factor:
\begin{equation}
 {\cal G}=\frac{{\cal B}^2\cal DE}{{\cal A}^2\cal C}.
\label{vs.corr1}
\end{equation}

(ii) The energy generation equation:
\begin{equation}
 \der Fz=Q_{dis}
\label{vs.P}
\end{equation}
where $F$ is the flux of energy generated inside the disk at a
rate given by $Q_{dis}$ - the dissipation rate which will be
discussed in \S \ref{sect.dissprof}.

(iii) The generated flux of energy is transported outward
through diffusion of radiation and/or convection according
to the following thermodynamical relation:
\begin{equation}
\nabla=\derln TP
\label{vs.gradient}
\end{equation}
where $\nabla$ is the thermodynamical gradient which can be either
radiative or convective:
\begin{equation}
\nabla=\left\{\begin{array}{lll}
\nabla_{rad} & {\rm dla} & \nabla_{rad}\le\nabla_{ad} \\
\nabla_{conv} & {\rm dla} & \nabla_{rad}>\nabla_{ad} \\
\end{array}\right.
\label{vs.grads}
\end{equation}
The radiative gradient $\nabla_{rad}$ is calculated using diffusive approximation:
\begin{equation}
\nabla_{rad}=\frac{3\kappa_R PF}{16\sigma T^4\Omega_K^2{\cal G}z}
\label{vs.gradrad}
\end{equation}
where $\kappa_R$ is the mean Rosseland opacity (see
\S\ref{sect.opacities}) and $\sigma$ is the Stefan-Boltzmann
radiation constant.

(iii) When the temperature gradient is steep enough to exceed the
value of the adiabatic gradient $\nabla_{ad}$ we have to consider
the convective energy flux. The convective gradient
$\nabla_{conv}$ is calculated using the mixing length theory
introduced by \citet{paczynski69}. Herein we take the following
mixing length:
\begin{equation}
H_{ml}=1.0 H_P
\label{vs.hml}
\end{equation}
with pressure scale height $H_P$ defined as \citep{hameury98}:
\begin{equation}
H_{P}=\frac P{\rho\Omega_K^2 z+\sqrt{P\rho}\Omega_K}
\label{vs.hP}
\end{equation}
The convective gradient is defined by the following formula:
\begin{equation}
\nabla_{conv}=\nabla_{conv}+(\nabla_{rad}-\nabla_{ad})y(y+V)
\label{vs.gradconv}
\end{equation}
where $y$ is the solution of the equation:
\begin{equation}
\frac94\frac{\tau_{ml}^2}{3+\tau_{ml}^2}y^3+Vy^2+V^2y-V=0
\label{vs.eqy1}
\end{equation}
with the typical optical depth for convection
$\tau_{ml}=\kappa_R\rho H_{ml}$ and $V$ given by:
\begin{equation}
\frac{1}{V^2}=\left(\frac{3+\tau_{ml}^2}{3\tau_{ml}}\right)^2
\frac{\Omega_K^2zH_{ml}^2\rho^2C_P^2}{512\sigma^2T^6H_P}
\left(\pderln\rho T\right)_P
\left(\nabla_{rad}-\nabla_{ad}\right)
\label{vs.eqy2}
\end{equation}
The thermodynamical quantities $C_P$, $\nabla_{ad}$ and $(\partial\,{\rm ln}\,\rho/\partial\,{\rm ln}\,T)_P$ are calculated using
standard formulae \citep[e.g.][]{stellarstructure} assuming solar abundances
($X=0.70$, $Y=0.28$) and taking into account, when necessary, the
effect of partial ionization of gas on the gas mean molecular
weight.

(iv) To close the set of equations describing the vertical structure
of an accretion disk we have to provide boundary conditions.
At the equatorial plane ($z=0$) we put $F=0$ while at the
disk photosphere we require $F=\sigma T^4$ with the flux $F$ 
given by \citet{pagethorne}.

\subsection{Opacities}
\label{sect.opacities}
In this work we consider optically thick accretion disks.
Therefore, we use Rosseland mean opacities $\kappa_R$. The
opacities as a function of density and temperature are taken from
\citet{alexander} and \citet{seaton}. For temperatures and
densities out of both domains we interpolate opacities between
these two tables for intermediate values.

\section{Numerical method}

The set of ordinary differential equations defined in
\S\ref{sect.equations} is solved using Runge-Kutta method of the
4th order. We start integrating from the equatorial plane ($z=0$)
where we assume some arbitrary central temperature $T_C$,
density $\rho_C$ and set $F=0$. We integrate given set of
equations until we reach the photosphere which is defined as a
layer where the following equation (corresponding to the optical
depth $\tau=2/3$):
\begin{equation}
\kappa_R P=\frac23 \Omega_K^2{\cal G} z
\label{nm.phot}
\end{equation}
is satisfied. If the temperature $T$ and flux $F$ does not satisfy
the outer boundary condition ($F=\sigma T^4$) we tune
up the value of the assumed central density $\rho_C$ and again
integrate starting from the equatorial plane. The converged
solution is usually obtained after a few iteration steps. To get
the photosphere profile $h(r)$ we look for the disk solutions at a
number of radii in the range $r_{ms}<r\le 200 M$ with $r_{ms}$
being the radius of the marginally stable orbit. At each radius we
additionally search for the value of the central temperature which
determines the solution with the emitted flux equal to the Novikov
\& Thorne value.


\section{The photosphere location for a few ad hoc assumed
 vertical dissipation profiles}


Herein we apply our simplified model of vertical structure
(\S\ref{sect.model}) to calculate photosphere location for
different dissipation prescriptions. We test four families of
dissipation functions described in the following section.
\subsection{Dissipation profiles}
\label{sect.dissprof}
First of all we apply the standard \textit{$\alpha P$
prescription} given by \citet{shakura-73}. The authors assumed
that the source of viscosity in accretion disk is connected with
turbulence in gas-dynamical flow the nature of which was unknown
at that time. They based on the following expression for the
kinematic viscosity coefficient $\nu$:
\begin{equation}
 \nu\approx v_{turb}l_{turb}
\label{dp.nu}
\end{equation}
(where $v_{turb}$ and $l_{turb}$ stand for typical turbulent
motion velocity and length scale, respectively). Shakura \&
Sunyaev assumed that the velocity of turbulent elements cannot
exceed the speed of sound $c_s$ as well as their typical size
cannot be larger that the disk thickness $h$. Taking into account
the expression for the $t_{r\phi}$ component of the viscous stress
tensor in a Newtonian accretion disk:
\begin{equation}
 t_{r\phi}=-\frac{3}{2}\rho\nu\Omega
\label{dp.trphi}
\end{equation}
and the standard form of the vertical equilibrium:
\begin{equation}
h\approx\left(\frac{P}{\rho}\right)^{1/2}\left(\frac{r^3}{GM}
\right)^{1/2}\approx\frac{c_s}{\Omega}
\label{dp.verteq}
\end{equation}
one can limit the $t_{r\phi}$ in the following way:
\begin{equation}
 -t_{r\phi}\leq\rho\Omega c_s h\approx \rho c_s^2\approx P
\label{dp.trphiest}
\end{equation}
Therefore, we may introduce a dimensionless viscosity parameter
$\alpha$ satisfying the condition $\alpha\leq 1$:
\begin{equation}
 t_{r\phi}=-\alpha P
\label{dp.alphaP}
\end{equation}
This expression for viscosity is called \textit{the $\alpha$
prescription} and has been widely and successfully used for many
years in accretion disk theory. In this case the flux generation
rate ${\rm d}F/{\rm d}z$ (Eq. \ref{vs.P}), given in the local rest
frame by:
\begin{equation}
\der Fz=\sigma_{(r)(\phi)}t_{(r)(\phi)}=
-\frac 32\frac{\cal DB}{\cal C}t_{(r)(\phi)}\Omega_K
\label{dp.aP.1}
\end{equation}
is expressed as:
\begin{equation}
Q_{dis}=\der Fz=\frac 32\frac{\cal DB}{\cal C}\alpha P\Omega_K
\label{dp.aP.2}
\end{equation}
In this work we apply the regular $\alpha P$ prescription for a
few arbitrary values of the $\alpha$ parameter: $\alpha = 0.001,
0.01, 0.1, 0.2, 0.5, 1.0$.

In the last decade several authors
\citep[e.g.][]{sincellkrolik97,hubenynonlte,davisblaes04,hui05}
used to apply a \textit{constant dissipation rate} per unit mass.
We follow them investigating a class of models using the following
formula for the flux generation rate:
\begin{equation}
Q_{dis}=\der Fz=\frac{F_0}{m_0}\rho
\label{dp.const.10}
\end{equation}
Due to the fact that this expression requires not only the total
emitted flux $F_0$ but also the total surface density $m_0$ one
has to provide the latter basing on other disk solutions. In our
work we use models calculated using regular $\alpha$ prescription
for different values of $\alpha$.

Recent numerical MHD simulations of stratified accretion disks in
shearing box approximation \citep{davisblaes09} show
that dissipation rate can be well approximated by the following
formula:
\begin{equation}
Q_{dis}=\der Fz=\frac12\frac{F_0}{\sqrt{m_0}}\frac1{\sqrt{m}}\rho
\label{dp.sig.10}
\end{equation}
We apply it in another class of models (\textit{$m^{-1/2}$
profile}). Similarly like in the \textit{constant dissipation
rate} case we impose surface density profile obtained using
$\alpha P$ models.

Several authors investigated accretion disks with heat generation
concentrated near the equatorial plane or close to the disk
surface \citep[e.g.][]{paczynskiheatinterior,
paczynskisurfaceaccretion}. Therefore, we implement one more class
of dissipation prescriptions with \textit{arbitrary positioned}
dissipation maximum. We use the following expression
\begin{equation}
Q_{dis}=\der Fz=\left.\der Fz\right|_{\alpha P}
e^{-\frac{(z-z_0)^2}{2d^2}}=\frac 32\frac{\cal DB}{\cal C}
\alpha P\Omega_Ke^{-\frac{(z-z_0)^2}{2d^2}}
\label{dp.arb.10}
\end{equation}
being flux generation rate for $\alpha P$ models re-scaled with
Gaussian function centered at $z_0$ with dispersion parameter $d$
given by:
\begin{equation}
z_0/r=d_{max}\quad\quad d/r=d_{disp}
\label{dp.arb.11}
\end{equation}
where $d_{max}$ and $d_{disp}$ are constant for a given disk
model.

\begin{figure}
  \resizebox{\hsize}{!}{\includegraphics{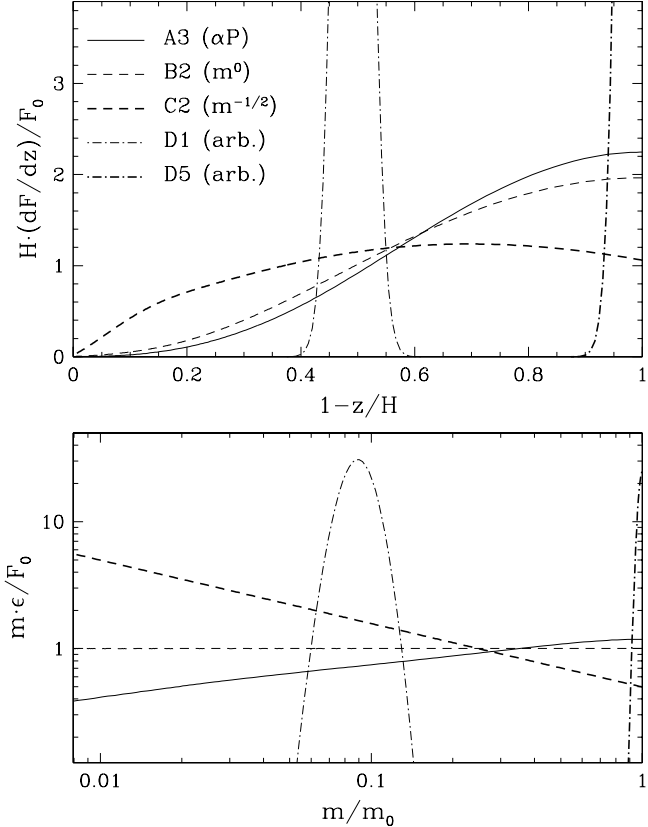}}
\caption{ Dissipation profiles for a few representative models at
$r=20M$. The upper panel presents dissipation functions versus the
vertical coordinate $z$ while the bottom panel versus surface
density $m(z)=\int_{H}^{z}\rho(z)\,dz'$. On both panels left and
right edges correspond to disk surface and equatorial plane,
respectively. The models represent four families: $aP$ (A3, solid line),
constant dissipation rate per unit mass (B2, thin dashed line),
$m^{-1/2}$ profile (C2, thick dashed line) and arbitrary
positioned dissipation function (D1 and D5, dash-dotted lines). }
  \label{f.qdis}
\end{figure}

In Fig. \ref{f.qdis} we present dissipation profiles for a few
exemplary models. The upper panel presents flux generation rate
versus the vertical coordinate. The $\alpha P$ model (solid line)
has bell-shaped dissipation profile where most of energy is
generated at $z<0.5 H$. The model with constant dissipation rate
per unit mass (thin dashed line) exhibits very similar behavior.
The dissipation in no longer concentrated around the equatorial
plane for the $m^{-1/2}$ model (thick dashed line) - the maximal
flux generation is located at $z=0.4H$. However, the dissipation
profile is still extended through the whole disk thickness. The
arbitrary positioned models (dot-dashed lines) show opposite
behavior: the energy generation is concentrated around an
arbitrary given location. In case of the models presented in Fig.
\ref{f.qdis} the dissipation is confined to altitudes close to the
equatorial plane ($z=0$) and $z=0.5H$.

Another point of view on the dissipation profiles is presented in
the bottom panel of Fig. \ref{f.qdis}. The profiles are plotted
against the fractional mass surface density $m/m_0$. The right
edge of the figure ($m=m_0$) corresponds to the equatorial plane.
Two families of models, with constant and proportional to
$m^{-1/2}$ dissipation profiles, are represented by straight lines
(they are power-law functions of $m$). As on the upper plot, one
can notice similarity between the $\alpha P$ and constant
dissipation rate models. The maximum of the flux generation rate
for the arbitrary positioned model centered at $z=0.5H$
corresponds to the mass depth $m=0.1m_0$. Obviously, the model
centered at the equatorial plane has maximal dissipation rate at
$m=m_0$.
\subsection{The photosphere location}
\label{s.photosphere}
\begin{figure}
  \resizebox{\hsize}{!}{\includegraphics{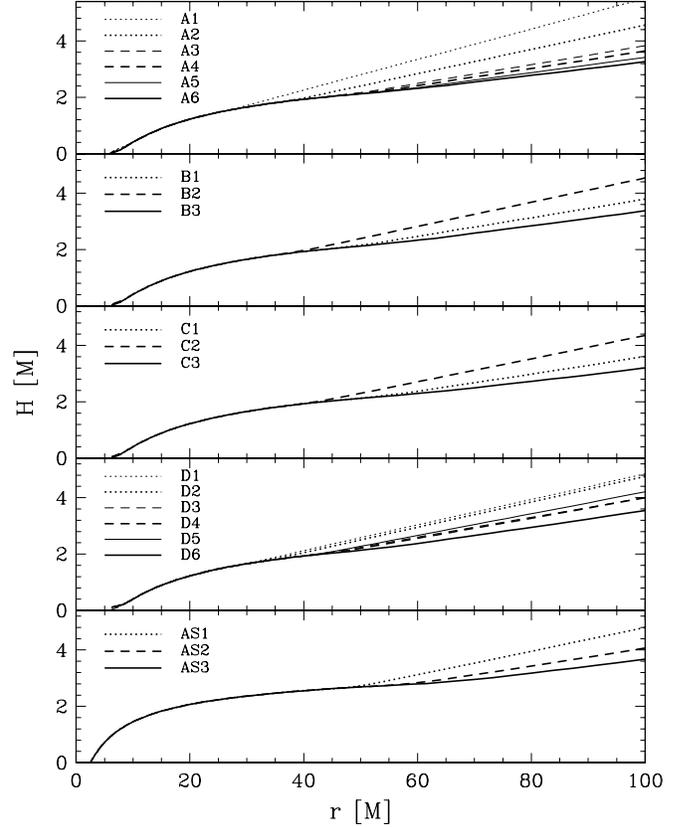}}
\caption{ Photosphere profiles for different model families for
$\dot m=0.1\dot m_C$. The uppermost panel presents profiles for
the $\alpha P$ models. The second one for models with constant
dissipation rate per unit mass. The third one for $m^{-1/2}$
dissipation profile. The forth is for arbitrary positioned
dissipation while the bottommost is for $\alpha P$ models in case
of a rotating BH ($a^*=0.9$). BH mass was $9.4\msun$. Model
parameters are given in Table \ref{t.models}. }
  \label{f.phot}
\end{figure}

We test a number of models representing all four classes of
dissipation profiles for a moderate mass accretion rate $\dot
m=0.1\dot m_C$. Details of model parameters are given in Tab.
\ref{t.models}. The location of the disk photospheres is presented
in Fig. \ref{f.phot}. The first four panels present results for
four classes of dissipation profiles in case of a non rotating BH.
Common behaviour is clearly visible: all photosphere locations
coincide at radii lower than $30M$. Outside this radius the
location of the photosphere depends on the assumed dissipation
profile and vary between $H/r\approx0.03$ for the $\alpha P$ model
with the highest value of $\alpha=1.0$ and $H/r\approx0.06$ for
the lowest value $\alpha=0.001$. It is of major importance to
understand why all photosphere profiles coincide in the inner
region which corresponds to the radiation pressure and electron
scattering dominated regime of an accretion disk. The location of
the photosphere is defined by Eq. \ref{nm.phot}. Wherever the
radiation pressure and electron scattering dominate in a disk the
left hand side of that formula depends only on the temperature
which is connected at the photosphere with the outcoming flux by
$F=\sigma T^4$. Therefore, the formula for the energy flux
(Eq. \ref{flux-kerr}) determines uniquely the location of the
photosphere in the inner region of a disk \textit{independently}
of the assumed dissipation function. Under these assumptions we
get from Eq. \ref{nm.phot}:
\be
H=\frac{2F\kappa_{es}}{\Omega^2_K{\cal G} c} \label{ph.H}
\ee
which describes perfectly the
photosphere profiles presented in Fig. \ref{f.phot} for $r<30M$.

\begin{figure}
  \resizebox{\hsize}{!}{\includegraphics{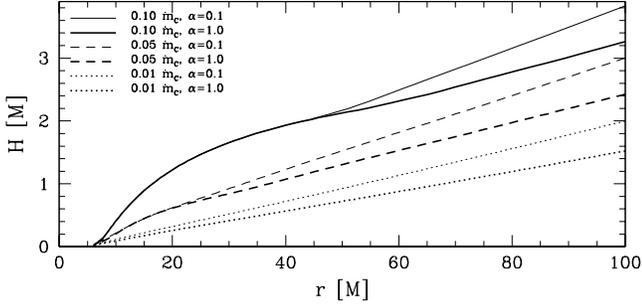}}
\caption{
Photosphere profiles for $\alpha P$ models for three different
mass accretion rates ($\dot m=0.10$, $0.05$ and $0.01
\dot m_C$) and two values of the $\alpha$ parameter ($0.1$
and $1.0$). The regions where the photosphere profiles
for a given accretion rate coincide are radiation pressure
dominated. BH was non-rotating with $M=9.4\msun$.
}
  \label{f.photmdot}
\end{figure}

Disk thickness profiles for a case of a rotating BH for three
$\alpha P$ models are presented in the bottommost panel. They
exhibit very similar behaviour: the photosphere location does not
depend on the viscosity prescription for small radii corresponding
to radiation pressure and electron scattering dominated region
which, in case of a rotating BH with $a^*=0.9$, extends inside
$r=40M$. Therefore, we may infer that such behaviour is general in
accretion disks which exhibit radiation pressure dominated
regimes. This statement is of paramount importance due to the fact
that most of the emission from an accretion disk take place (in
non-rotating case) between $6$ and $20M$. As we have proven, for
such radii the photosphere location should not depend on assumed
dissipation profile and, therefore, the spectrum is likely to be
independent of dissipation profile, either. However, one cannot
make sure without performing full ray-tracing which would account
for flux emitted from dissipation dependent regions correctly.
Results of this procedure will be discussed in the following
section.
\begin{table}
 \caption{Model Assumptions for $\dot m=0.1\dot m_C$}
\label{t.models}

\begin{tabular}{llcccc}
\hline\hline
Model & Model family\tablenotemark{a}                 & $\alpha$             & $d_{max}$     & $d_{disp}$ & $a^*$   \\
\hline
A1    & $\alpha P$                                    & 0.001                & -             & -          & 0.0          \\
A2    & $\alpha P$                                    & 0.01                 & -             & -          & 0.0          \\
A3    & $\alpha P$                                    & 0.1                  & -             & -          & 0.0          \\
A4    & $\alpha P$                                    & 0.2                  & -             & -          & 0.0          \\
A5    & $\alpha P$                                    & 0.5                  & -             & -          & 0.0          \\
A6    & $\alpha P$                                    & 1.0                  & -             & -          & 0.0          \\
AS1   & $\alpha P$                                    & 0.01                 & -             & -          & 0.9          \\
AS2   & $\alpha P$                                    & 0.1                  & -             & -          & 0.9          \\
AS3   & $\alpha P$                                    & 0.5                  & -             & -          & 0.9          \\
B1    & constant diss. rate& 0.01\tablenotemark{b}& -             & -          & 0.0          \\
B2    & constant diss. rate& 0.1\tablenotemark{b} & -             & -          & 0.0          \\
B3    & constant diss. rate& 0.5\tablenotemark{b} & -             & -          & 0.0          \\
C1    & $m^{-1/2}$ & 0.01\tablenotemark{b}& -             & -          & 0.0          \\
C2    & $m^{-1/2}$ & 0.1\tablenotemark{b} & -             & -          & 0.0          \\
C3    & $m^{-1/2}$ & 0.5\tablenotemark{b} & -             & -          & 0.0          \\
D1    & arbitrary positioned             & 0.1                  & 0.0           & 0.001      & 0.0          \\
D2    & arbitrary positioned             & 0.1                  & 0.02          & 0.001      & 0.0          \\
D3    & arbitrary positioned             & 1.0                  & 0.0           & 0.001      & 0.0          \\
D4    & arbitrary positioned             & 1.0                  & 0.02          & 0.001      & 0.0          \\
D5    & arbitrary positioned             & 1.0                  & 0.02          & 0.0005     & 0.0          \\
D6   & arbitrary positioned             & 1.0                  & 0.02          & 0.005      & 0.0          \\
\hline
\tablenotetext{a}{Details of model assumptions are given in \S\ref{sect.dissprof}}
\tablenotetext{b}{Calculated basing on surface density profiles obtained in the $\alpha P$ model with given $\alpha$ and $a^*$}
\end{tabular}
\end{table}
As \cite{shakura-73} have shown for the lowest mass accretion
rates the gas dominated region of an accretion disk may extend all
the way down to the inner edge of a disk. For such case the
behavior described in the previous paragraph is not expected: the
photosphere profile should depend on the dissipation function at
all radii. To account for this fact we compare photosphere
profiles of accretion disks with low mass accretion rates (Fig.
\ref{f.photmdot}). For $\dot m=0.1\dot m_C$ the radiation
dominated region extends upto $r\approx40M$. For lower accretion
rate ($0.05\dot m_C$) it is confined to $r<20M$, while for the
lowest ($0.01\dot m_C$) it never appears. In that case the
photosphere location indeed depends on the assumed viscosity
prescription even for radii lower than $20M$ where most of the
emission comes out. However, one has to bear in mind that $H/r$
ratio for such low mass accretion rate is even below $0.015$ and
the impact of the photosphere location on the emerging spectrum
should be insubstantial. It is studied in details in the following
section.
\subsection{Spectra}

\begin{figure}
  \resizebox{\hsize}{!}{\includegraphics{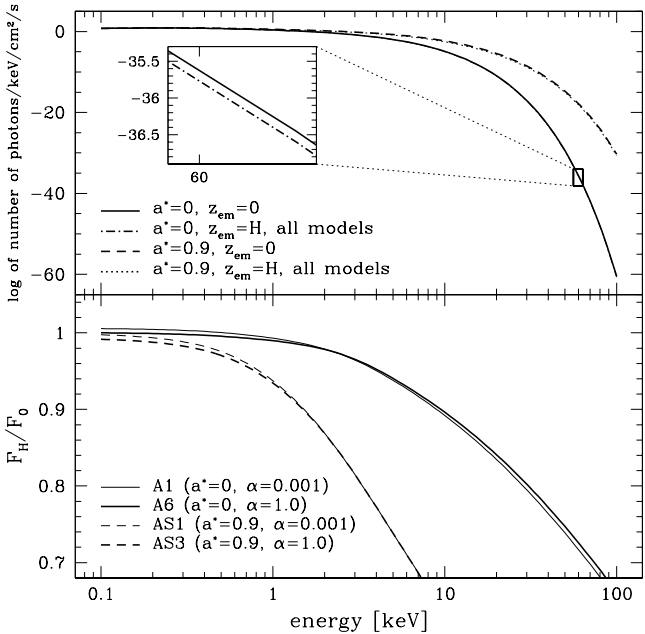}}
\caption{ Emerging spectra calculated for different models. The
upper panel presents number of photons per energy crossing every
second a unit surface assuming $d=10\rm kpc$. Spectra calculated
assuming emission from the equatorial plane ($z_{em}=0$) as well
as from the photosphere surface ($z_{em}=H$) are presented for
non-rotating ($a^*=0$) and highly-spinning ($a^*=0.9$) BH. The
spectra for different dissipation profiles are indistinguishable -
they all coincide with dot-dashed and dotted lines (depending on
BH spin). The bottom panel presents ratio of the spectral flux
(defined as above) calculated from the photosphere to the flux
calculated from the equatorial plane for the models with extreme
photosphere profiles (see \S\ref{s.photosphere}). The inclination
angle $i=60^\circ$. }
  \label{f.spectra}
\end{figure}

\begin{figure}
  \resizebox{\hsize}{!}{\includegraphics{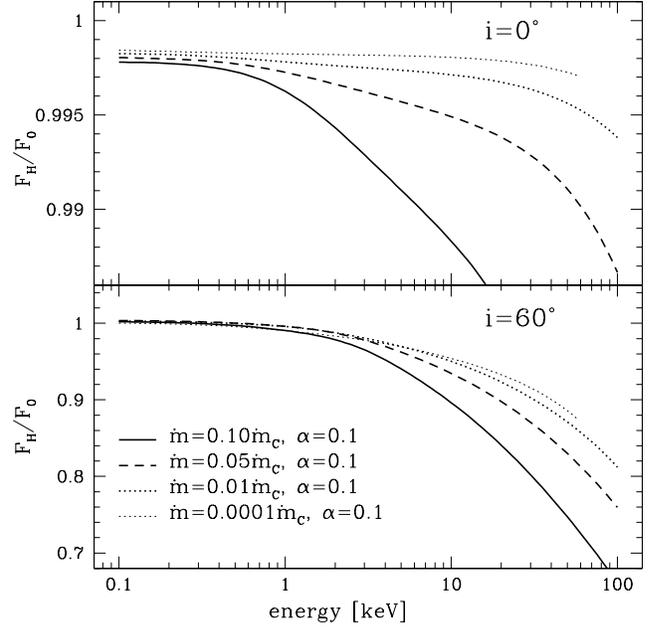}}
\caption{
Ratios of spectral profiles (photosphere to equatorial)
for models with low mass accretion rates for the perpendicular
($i=0^\circ$, upper panel) and $i=60^\circ$ (lower panel) observers.
}
  \label{f.spectraratiomdot}
\end{figure}

To assess the importance of different vertical dissipation
profiles on emerging spectra we calculate them basing on
photosphere profiles obtained using our model. For comparison, we
also account for spectra calculated assuming emission from the
equatorial plane. Spectra for models with mass accretion rate
$\dot m=0.1\dot m_C$ are presented in Fig. \ref{f.spectra}. The
upper panel presents total spectra for non- and rapidly-rotating
BH. The spectral profile for $a^*=0.9$ case extends to higher
energies due to the fact that accretion disk inner edge moves
closer to the central object with increasing BH angular momentum
extracting more energetic photons. Spectra for all the photosphere
profiles almost coincide and differ significantly from spectra
calculated assuming emission from the equatorial plane. It is
clearly visible on the lower panel where we plot ratio of spectral
fluxes emitted from the photosphere to emitted from the equatorial
plane for two most extreme photosphere profiles (compare Fig.
\ref{f.phot}) for each value of BH angular momentum. The spectral
profiles are identical up to a factor of $1\%$. The agreement is
the best for energies of few $\rm keV$. One may conclude that for
accretion disks with sufficiently high accretion rate to form the
inner radiation pressure supported region, the spectra depend 
only insignificantly on vertical dissipation profile in the most
interesting range of energies
\citep[$2.5\div20\rm\,keV$;][]{lmcx1spin}.
As was discussed in \S\ref{s.photosphere} for the lowest mass
accretion rates ($<0.05\dot m_C$) the inner, radiation pressure
dominated disk region never appears and the photosphere location
depends on the dissipation profile down to the disk inner edge. As
it has been pointed out, such regime of accretion rates implies
small values of the $H/R$ ratio. To assess the impact of the
dissipation profiles on the spectra we plotted ratios of
corresponding spectral profiles for low and very low mass
accretion rates (Fig. \ref{f.spectraratiomdot}). The upper panel
presents the ratios as observed by an observer perpendicular to
the disk plane. The ratios approach $1$ with decreasing mass
accretion rate. The convergence is very slow due to the fact that
disk thickness in disk outer region (gas pressure and free-free
scaterings dominated) depends weakly on the mass accretion rate
\citep[$H\sim \dot m^{3/20}$, ][]{shakura-73}. The difference in
spectral fluxes between models with $\dot m=0.1\dot m_C$ and
$0.0001\dot m_C$ is no larger than $0.1\%$ at $20\rm keV$. For
inclined observer (the bottom panel) the departures from the
equatorial plane model are much more significant (up to $10\%$ at
$20\rm keV$ even for the lowest mass accretion rate). However, the
difference between the spectral fluxes for the models with the two
lowest mass accretion rates (corresponding to $H/R=0.007$ and
$0.015$, respectively) is still of the order of $1\%$ or even
smaller. We conclude that in the lowest accretion rate regime
changes of the photosphere location caused either by different
mass accretion rates or different dissipation profiles are
insignificant. However, one has to keep in mind that even for mass 
accretion rates as small as $0.0001\dot m_C$ taking into account
the non-zero photosphere location is important for inclined
observers as the spectral flux at high energies approaches the
equatorial plane limit very slowly.


\section{Discussion}


Results presented here are very encouraging in the context of
fitting the calculated optically thick spectra of slim accretion
BH disk to the observed spectra. Our results imply that the major
uncertainties of the accretion disk models, in particular the
vertical dissipation profiles, have a rather small influence on
the calculated spectra of steady disks with low accretion rates.
This is a good news for these who use spectral fitting to estimate
the black hole spin.

One should have in mind, however, that there are several effects
that have been not taken into account in our simple model, and
more theoretical work is needed before fully believable methods of
calculating slim disk spectra will be at hand. One of the most
obvious (and also quite challenging) theoretical developments
needed is a self-consistent and simultaneous  solving the vertical
and radial structures of slim, optically thick black hole
accretion disk. We are working on this problem.
\begin{acknowledgements} 
This work was initiated at the workshop "Turbulence and Oscillations in
Accretion Discs" held 1-15 October 2008 at Nordita in Stockholm. We
acknowledge Nordita's support. We thank Omer Blaes for his invaluable support and helpful comments.
We also acknowledge support from
Polish Ministry of Science grants
N203 0093/1466 and 3040/B/H03/2008/35,
Swedish Research Council grant
VR Dnr 621-2006-3288 and Czech Ministry of Science grant GAAV 300030510.
\end{acknowledgements}

\bibliographystyle{aa}
\bibliography{photosphere}
%


\end{document}